\documentclass[usenatbib]{mn2e}
\usepackage{graphicx,natbib}
\usepackage{bm,url,color}
\graphicspath{{./fig/}{./png/}}

\newcommand{\EQ}{\begin{equation}}
\newcommand{\EN}{\end{equation}}
\newcommand{\EQA}{\begin{eqnarray}}
\newcommand{\ENA}{\end{eqnarray}}
\newcommand{\eq}[1]{(\ref{#1})}

\newcommand{\Eq}[1]{Equation~(\ref{#1})}

\newcommand{\EqsS}[2]{Equations~(\ref{#1}), (\ref{#2})}

\newcommand{\Sec}[1]{Sect.~\ref{#1}}

\newcommand{\Fig}[1]{Fig.~\ref{#1}}

\newcommand{\Tab}[1]{Table~\ref{#1}}

\newcommand{\bra}[1]{\langle #1\rangle}

{}
{}
{}

{}
{}
{}
{}
{}
{}
{}
{}
{}
{}
{}
{}
{}
{}
{}
{}
{}

{}

\newcommand{\hatkk}{\hat{\bm{k}}}

{}

{}
{}
{}

\newcommand{\kk}{\bm{k}}

\newcommand{\xx}{\bm{x}}

\newcommand{\BB}{\bm{B}}

\newcommand{\JJ}{\bm{J}}

\newcommand{\AAA}{\bm{A}}

\newcommand{\uu}{\bm{u}}

\newcommand{\ee}{\mbox{\boldmath $e$} {}}

\newcommand{\ff}{\mbox{\boldmath $f$} {}}

\newcommand{\nab}{{\bm{\nabla}}}

\newcommand{\SSSS}{\mbox{\boldmath ${\sf S}$} {}}

\newcommand{\ii}{{\rm i}}

\newcommand{\SKEW}{{\rm skew}}

\newcommand{\DD}{{\rm D} {}}

\newcommand{\dd}{{\rm d} {}}

\def\ga{\mathrel{\mathchoice {\vcenter{\offinterlineskip\halign{\hfil
$\displaystyle##$\hfil\cr>\cr\sim\cr}}}
{\vcenter{\offinterlineskip\halign{\hfil$\textstyle##$\hfil\cr>\cr\sim\cr}}}
{\vcenter{\offinterlineskip\halign{\hfil$\scriptstyle##$\hfil\cr>\cr\sim\cr}}}
{\vcenter{\offinterlineskip\halign{\hfil$\scriptscriptstyle##$\hfil\cr>\cr\sim\cr}}}}}
\def\St{\mbox{\rm St}}
\def\tauAD{\tau_{\rm AD}}
\def\StAD{\mbox{\rm St}_{\rm AD}}

\def\PAD{\mbox{\rm Pr}_{\rm AD}}
\def\Pm{\mbox{\rm Pr}_{\rm M}}
\def\Rm{\mbox{\rm Re}_{\rm M}}
\def\Rmc{\mbox{\rm Re}_{\rm M,c}}
\def\Rmz{\mbox{\rm Re}_{\rm M0}}

\def\EEM{{\cal E}_{\rm M}}

\def\Ei{E_{\rm i}}

\def\EK{E_{\rm K}}
\def\EM{E_{\rm M}}

\def\cs{c_{\rm s}}

\def\vA{v_{\rm A}}

\def\kf{k_{\rm f}}

\def\EM{E_{\rm M}}

\def\epsK{\epsilon_{\rm K}}
\def\epsM{\epsilon_{\rm M}}
\def\epsAD{\epsilon_{\rm AD}}

\def\Brms{B_{\rm rms}}

\def\urms{u_{\rm rms}}
\def\urmsz{u_{\rm rms0}}

\def\etaAD{\eta_{\rm AD}}

\def\Beq{B_{\rm eq}}

\def\half{{\textstyle{1\over2}}}

\def\onethird{{\textstyle{1\over3}}}

\newcommand{\s}{\,{\rm s}}

\newcommand{\CM}{{\rm cm}}
\newcommand{\cm}{\,{\rm cm}}

\newcommand{\km}{\,{\rm km}}

\newcommand{\pc}{\,{\rm pc}}

\newcommand{\yapj}[3]{ #1, {ApJ,} {#2}, #3}

\newcommand{\yapjl}[3]{ #1, {ApJ,} {#2}, #3}

\newcommand{\yana}[3]{ #1, {A\&A,} {#2}, #3}

\newcommand{\yprl}[3]{ #1, {Phys.\ Rev.\ Lett.,} {#2}, #3}

\newcommand{\ymn}[3]{ #1, {MNRAS,} {#2}, #3}
\newcommand{\ynat}[3]{ #1, {Nature,} {#2}, #3}

\newcommand{\ypre}[3]{ #1, {Phys.\ Rev.\ E,} {#2}, #3}
\newcommand{\ypnas}[3]{ #1, {Proc.\ Nat.\ Acad.\ Sci.,} {#2}, #3}

\newcommand{\yspd}[3]{ #1, {Sov.\ Phys.\ Dokl.,} {#2}, #3}
\newcommand{\yjcp}[3]{ #1, {J.\ Comput.\ Phys.,} {#2}, #3}
\newcommand{\yjour}[4]{ #1, {#2}, {#3}, #4}

\newcommand{\yproc}[5]{ #1, in {#3}, ed.\ #4 (#5), #2}

\newcommand{\papj}[2]{ #1, {ApJ}, in press, arXiv:#2}

\title{Ambipolar diffusion in large Prandtl number turbulence}
\author[Axel Brandenburg]{
Axel Brandenburg$^{1,2,3,4}$\thanks{E-mail:brandenb@nordita.org}
\\
$^1$Nordita, KTH Royal Institute of Technology and Stockholm University, Roslagstullsbacken 23, SE-10691 Stockholm, Sweden\\
$^2$Department of Astronomy, AlbaNova University Center, Stockholm University, SE-10691 Stockholm, Sweden\\
$^3$JILA and Laboratory for Atmospheric and Space Physics, University of Colorado, Boulder, CO 80303, USA\\
$^4$McWilliams Center for Cosmology \& Department of Physics, Carnegie Mellon University, Pittsburgh, PA 15213, USA
}

\date{Accepted 2019 May 28. Received 2019 May 23; in original form 2019 March 21}
\begin{document}
\maketitle

\begin{abstract}
We study the effects of ambipolar diffusion (AD) on hydromagnetic
turbulence.
We consider the regime of large magnetic Prandtl number, relevant to
the interstellar medium.
In most of the cases, we use the single fluid approximation where
the drift velocity between charged and neutral particles is proportional
to the Lorentz force.
In two cases we also compare with the corresponding two-fluid model,
where ionization and recombination are included in the continuity and
momentum equations for the neutral and charged species.
The magnetic field properties are found to be well represented by
the single fluid approximation.
We quantify the effects of AD on total and spectral kinetic and magnetic
energies, the Ohmic and AD dissipation rates, the statistics of the
magnetic field, the current density, and the linear polarization as
measured by the rotationally invariant $E$ and $B$ mode polarizations.
We show that the kurtosis of the magnetic field decreases with
increasing AD.
The $E$ mode polarization changes its skewness from positive
values for small AD to negative ones for large AD.
Even when AD is weak, changes in AD have a marked effect on the skewness
and kurtosis of $E$, and only a weak effect on those of $B$.
These results open the possibility of employing $E$ and $B$ mode
polarizations as diagnostic tools for characterizing turbulent properties
of the interstellar medium.
\end{abstract}

\begin{keywords}
turbulence --- ISM
\end{keywords}

\section{Introduction}

In the cool parts of the interstellar medium (ISM), the ionization
fraction is low, so ions and neutrals move at different speeds, whose
difference is given by the ambipolar diffusion (AD) speed.
Particularly insightful is the single fluid model in the strong coupling
approximation for cases with negligible electron pressure.
It is then easy to see that there is not only enhanced diffusion, but
there is also a contribution to the electromotive force proportional
to the magnetic field, akin to the $\alpha$ effect in mean-field
electrodynamics.
Both terms increase with increasing magnetic field strength, making the
problem highly nonlinear.
In particular, AD can lead to the formation of sharp structures
\citep{BZ94}, an effect that has also been seen in the full two-fluid
description \citep{BZ95}.
It was already known for some time that, unlike Ohmic diffusion, AD
does not contribute to terminating the turbulent magnetic cascade,
even though both imply a removal of magnetic energy.
This became obvious when \cite{BS00} simulated the hydromagnetic forward
and inverse cascades in the presence of AD (see their Figure~2) to
understand its effect in the context of helical turbulent dynamos when
using it as a nonlinear closure, as was done by \cite{Sub99}.
The presence of magnetic helicity in this case made the interpretation
of the results more complicated, because the $\alpha$ effect-like
term of AD might then have been responsible for the apparent lack of
diffusive behavior.
For this reason, it is important to repeat similar calculations without
helicity, i.e., when there is only small-scale dynamo action.

The purpose of the present paper is to study AD in the context of a
small-scale dynamo, i.e., one that operates in non-helical homogeneous
turbulence.
Here, as discussed above, the $\alpha$ effect-like term proportional to
the magnetic field is expected to be negligible, because it involves
the current helicity density, and there is no reason for it to be of
significant magnitude when the turbulence is nonhelical.
It is therefore not obvious in which way AD affects the forward turbulent
cascade of kinetic and magnetic energies.

The problem of a nonhelical dynamo in the presence of AD has been
addressed by \cite{XL16} and \cite{Xu_etal19}.
They used a two-fluid description, which can have the advantage that no severe
(diffusive) time-step constraint occurs when the magnetic field reaches
saturation.
In their numerical work, \cite{Xu_etal19} focused on verifying the
linear growth during the damping stage of the dynamo near saturation,
which \cite{XL16} found in their earlier work.
However, ionization and recombination reactions are here neglected.
Those turn out to be important for allowing the formation of sharp
structures around magnetic nulls.
Recombination provides a sink for the charged species near magnetic nulls.
These species (ions and electrons) continue to concentrate the
field further, recombine at the null, and drift outward as neutrals
\citep{BZ95}.
This effect is important for alleviating an otherwise excessive electron
pressure near magnetic nulls, which would counteract the formation of
sharp structures.
We demonstrate the equivalence between the single fluid and the two-fluid
approaches in two particular cases that are of relevance to the present
paper.

For the purpose of the present work, we are particularly interested in
turbulent dynamos at large magnetic Prandtl numbers, which is relevant
for modelling the interstellar medium (ISM).
In this regime, the viscosity is large compared with the magnetic
diffusivity.
This leads to a truncation of the kinetic energy spectrum at a wavenumber
that is well below that of the magnetic energy; see the simulations of
\cite{HBD04} and \cite{Scheko04}.
In the ISM, the value of $\Pm$ is of the order of $10^{11}$ \citep{BS05},
but here we will only be able to simulate values of $\Pm$ of about a
few hundred.
Nevertheless, we may then already expect to see a clear effect on the
magnetic dissipative effects and, in particular, on the
kinetic to magnetic energy dissipation ratio, which is known to scale like
$\Pm^{0.3}$ when there is small-scale dynamo action, and like $\Pm^{0.7}$
when there is large-scale dynamo actions; see \cite{Bra14}.
It is a priori unclear how AD affects this dissipation ratio.
Again, within the strong coupling approximation, we would expect that
larger magnetic diffusion enhances the magnetic energy dissipation.
Naively, this would correspond to the case of a reduced effective value of
$\Pm$, so the effective value of the ratio $\epsK/\epsM$ should decrease.
Such a result might still be compatible with the usual $\Pm$ scaling
if $\Pm$ is interpreted as an effective magnetic Prandtl number that
would then also be reduced by AD.
It will then be interesting to see how the individual values of
$\epsK$ and $\epsM$ change.
In this context, it must be emphasized that in the statistically steady
state, $\epsM$ must be equal to the work done against the Lorentz force,
which corresponds to the rate of kinetic to magnetic energy conversion.
Therefore, a change in the dissipative properties both through ohmic
resistivity and through AD must also affect the kinetic to magnetic
energy conversion.
These questions will therefore also be clarified in the present work.

\section{The model}
\label{Model}

\subsection{The two-fluid description}
\label{2Fluid}

Before stating the governing equations in the single fluid
approximation, which will be adopted for most of the calculations
presented below, we first discuss the underlying two-fluid
equations for the neutral and ionized species \citep{Dra86}.
We emphasize that the ionized fluid component consists of ions and
electrons, both of which are assumed to be tightly coupled to each other.
We give the governing equations here in the form as used by \cite{BZ95},
\EQ
{\partial\AAA\over\partial t}=\uu_{\rm i}\times\BB-\eta\mu_0\JJ,
\label{dAdt1}
\EN
\EQ
\!\rho_{\rm i}{\DD\uu_{\rm i}\over\DD_{\rm i} t}=\JJ\times\BB-\nab p_{\rm i}
+\nab\cdot(2\nu\rho_{\rm i}\SSSS_{\rm i})
-\rho(\rho_{\rm i}\gamma+\zeta)(\uu_{\rm i}-\uu),\;
\label{DuiDt}
\EN
\EQ
\rho{\DD\uu\over\DD t}=\rho\ff-\nab p
+\nab\cdot(2\nu\rho\SSSS)
+\rho_{\rm i}(\rho\gamma+\alpha\rho_{\rm i})(\uu_{\rm i}-\uu),
\label{DuDt1}
\EN
\EQ
{\DD\ln\rho_{\rm i}\over\DD_{\rm i} t}=-\nab\cdot\uu_{\rm i}
+\zeta\rho/\rho_{\rm i}-\alpha\rho_{\rm i},
\EN
\EQ
{\DD\ln\rho\over\DD t}=-\nab\cdot\uu
-\zeta+\alpha\rho_{\rm i}^2/\rho,
\label{DlnrhoDt1}
\EN
where $\DD/\DD_{\rm i}t=\partial/\partial t+\uu_{\rm i}\cdot\nab$ and
$\DD/\DD t=\partial/\partial t+\uu\cdot\nab$ are the advection operators
for the ionized and neutral species, respectively,
$\uu_{\rm i}$ and $\uu$ are their velocities,
$\rho_{\rm i}$ and $\rho$ are their densities,
$p_{\rm i}$ and $p$ are their pressures,
$\zeta$ is the rate of ionization, $\alpha$ is the rate of recombination,
$\gamma$ is the drag coefficient between ionized and neutral fluids,
$\AAA$ is the magnetic vector potential,
$\BB=\nab\times\AAA$ is the magnetic field,
$\JJ=\nab\times\BB/\mu_0$ is the current density,
$\mu_0$ is the vacuum permeability,
${\sf S}_{ij}=\half(u_{i,j}+u_{j,i})-\onethird\delta_{ij}\nab\cdot\uu$
are the components of the traceless rate of strain tensor $\SSSS$,
with a roman subscript i in \Eq{DuiDt} denoting the analogous expression
for the ionized fluid, and
$\ff$ is a nonhelical monochromatic forcing function with wavevectors
$\kk(t)$ that change randomly at each time step and are taken from a
band of wavenumbers around a given forcing wavenumber $\kf$.
The forcing function is proportional to $\kk\times\ee$, where
$\ee$ is a random unit vector that is not parallel to $\kk$;
see \cite{HBD04} for details.
We adopt an isothermal equation of state with equal
and constant sound speeds $\cs$ for the ionized and
neutral components, such that their pressures are given by
$p_{\rm i}=\rho_{\rm i}\cs^2$ and $p=\rho\cs^2$, respectively.

\subsection{Single fluid approximation}
\label{Eqns}

In most of this work, we adopt the single fluid approximation, i.e.,
we assume that the electron pressure (which is equal to $p_{\rm i}$) can
be omitted and that the term $\rho\rho_{\rm i}\gamma(\uu_{\rm i}-\uu)$
in \Eq{DuiDt} is being balanced by $\JJ\times\BB$.
We can then replace $\uu_{\rm i}$ in \Eq{dAdt1} by
$\uu+\uu_{\rm AD}$, where $\uu_{\rm AD}=(\tauAD/\rho_0)\,\JJ\times\BB$
is the ambipolar drift velocity with $\tauAD=(\gamma\rho_{\rm i0})^{-1}$
being the mean neutral--ion collision time, and $\rho_{\rm i0}$ and $\rho_0$
are the initial density of ions and neutrals.
We thus solve the equations for $\AAA$, $\uu$, and $\rho$ in the form
\EQ
{\partial\AAA\over\partial t}=(\uu+\uu_{\rm AD})\times\BB-\eta\mu_0\JJ,
\label{dAdt}
\EN
\EQ
\rho{\DD\uu\over\DD t}=\rho\ff-\nab p+\nab\cdot(2\nu\rho\SSSS)+\JJ\times\BB,
\EN
\EQ
{\DD\ln\rho\over\DD t}=-\nab\cdot\uu.
\EN
As we demonstrate below, the solutions to these equations agree with
those to \EqsS{dAdt1}{DuDt1}, and \eq{DlnrhoDt1} when $\zeta$ and $\alpha$
are large enough (so that the electron pressure becomes negligible) and
$\gamma$ is large enough to ensure strong coupling between the ionized
and neutral fluids.

\subsection{Setup of the models and control parameters}

We consider a cubic domain of size $L^3$, so the smallest wavenumber
is $k_1=2\pi/L$.
We normally use the nominal average value $\kf=1.5\,k_1$, but, following
the reasoning of \cite{BHLS18}, we also use the effective value of $\kf$
that determines the relevant value of the magnetic Reynolds number,
\EQ
\Rm=\urms/\eta\kf^{\rm eff},
\EN
where $\kf^{\rm eff}\approx2\,k_1$ when $\kf=1.5\,k_1$.
This adjustment at the smallest wavenumber is motivated by the fact
that at such small wavenumbers, only 20 different vectors fall into
the wavenumber band with $|\kk|/k_1$ between $1$ and $2$, making this
a special case compared with those where $\kf$ is larger.

We normally evaluate $\Rm$ in saturated cases where the magnetic
field leads to a certain suppression of $\urms$.
In some cases, for example when specifying the critical growth rate
of the dynamo, it is advantageous to use instead the kinematic rms
velocity, $\urmsz$, and thus define $\Rmz=\urmsz/\eta\kf^{\rm eff}$.

The relative importance of viscous to magnetic diffusion is quantified
by the magnetic Prandtl number,
\EQ
\Pm=\nu/\eta.
\label{PrM_def}
\EN
For the single fluid models, we consider two types of runs,
one with $\Pm=20$ (series~I) and another with $\Pm=200$ (series~II).
In both cases, $\eta$ is unchanged and only $\nu$ is increased by a
factor of 10.
This implies that kinetic energy dissipation should occur at small
wavenumbers.
Our two-fluid models are similar to the single fluid models of
series~II.

We often express time scales in units of the sound travel time,
$\tau_{\rm s}=(\cs k_1)^{-1}$.
The correspondingly normalized quantities are denoted by a prime,
so we define
\EQ
\tauAD'\equiv\tauAD\cs k_1,\quad
\zeta'\equiv\zeta/\cs k_1,\quad \mbox{and}\quad
\gamma'\equiv\rho_0\gamma/\cs k_1.
\EN
Alternatively, we express $\tauAD$ in terms of the turbulent turnover
time $\tau_0=(\urmsz\kf)^{-1}$.
In particular, we define a generalized Strouhal number as
\EQ
\St_{\rm AD}=\tauAD\urmsz\kf\equiv\tauAD/\tau_0.
\label{StADdef}
\EN
We also define the quantity $k_{\rm AD}=\kf/\St_{\rm AD}$ as a
characteristic AD wavenumber where the turbulent and AD timescales
are comparable.
Note that we have used $\urmsz$ in the definition of $k_{\rm AD}$
instead of the actual rms velocity, which can be smaller by up to
a quarter when the magnetic field becomes strong and $\tauAD$ is not
too large.
Thus, the actual value of $k_{\rm AD}$ becomes reduced as the magnetic
field saturates.

For comparison with the cold interstellar medium, let us estimate
$\tauAD=n_{\rm n}/n_{\rm i}\nu_{\rm in}\approx7\times10^{14}\s$,
where we have used $n_{\rm n}=1\cm^{-3}$ and $n_{\rm i}\approx
1.1\times10^{-5}(n_{\rm n}/\CM^{-3})^{1/2}$ \citep{McKee}
for the neutral and ion number densities, and
$\nu_{\rm in}\approx1.3\times10^{-10}(n_{\rm n}/\CM^{-3})\s^{-1}$ \citep{Dra83}.
This gives $\tauAD'\approx7$ for $\cs=0.3\km\s^{-1}$ and
$k_1=1\pc^{-1}$.
Furthermore, using $\zeta=3\times10^{-17}$ to $10^{-15}\s^{-1}$ 
\citep{McCall}, we have $\zeta'=3\times10^{-3}$ to $0.1$.
The values of $\tauAD'$ and $\zeta'$ are comparable to those explored
below.

For our numerical simulations we use the
\textsc{Pencil Code}\footnote{\url{https://github.com/pencil-code},
DOI:10.5281/zenodo.2315093},
which is a high-order public domain code
for solving partial differential equations, including the hydromagnetic
equations given above.
It uses sixth order finite differences in space and the third order
2N-RK3 low storage Runge--Kutta time stepping scheme of \cite{Wil80}.
We use $576^3$ meshpoints for all runs in three dimensions and
$576$ meshpoints for our one-dimensional runs.

\subsection{Energy dissipation}
\label{EnergyDissipation}

For each of the two series, we vary the value of $\tauAD$ and
express it in terms of $\St_{\rm AD}$; see \Eq{StADdef}.
We also monitor the mean kinetic and magnetic energy dissipation
rates, $\epsK=\bra{2\nu\rho\SSSS^2}$ and $\epsM=\bra{\eta\mu_0\JJ^2}$,
respectively, where angle brackets denote volume averaging.
For Kolmogorov-type turbulence, the kinetic and magnetic
dissipation wavenumbers are given by $k_\nu=(\epsK/\nu^3)^{1/4}$ and
$k_\eta=(\epsM/\eta^3)^{1/4}$, respectively.

It is important to note that AD significantly adds to the rate of magnetic
energy dissipation \citep{Padoan_etal00,KC12}.
This becomes evident when looking at the magnetic energy equation,
\EQ
{\dd\EEM\over\dd t}=-W_{\rm Lor}-\epsAD-\epsM,
\EN
where $\EEM=\bra{\BB^2/2\mu_0}$ is the mean magnetic energy density
and $W_{\rm Lor}=\bra{\uu\cdot(\JJ\times\BB)}$ is the work done by the
Lorentz force.
The quantities $\epsAD=(\tau_{\rm AD}/\rho_0)\bra{(\JJ\times\BB)^2}$
and $\epsM=\bra{\eta\mu_0\JJ^2}$ are the loss terms corresponding to
AD and resistive heating, respectively.
In all cases presented here, we express the magnetic field strength in units
of the equipartition value $\Beq=\sqrt{\mu_0\rho_0}\,\urms$, which is being
evaluated during the saturation phase.
Given that AD contributes to magnetic energy dissipation, it will also be
important to define the resulting enhancement of the effective magnetic
diffusivity due to AD.
For this purpose, we rewrite part of the right-hand side of \Eq{dAdt} as
\EQ
\uu_{\rm AD}\times\BB-\eta\mu_0\JJ
=\alpha_{\rm AD}\BB-(\eta+\eta_{\rm AD})\mu_0\JJ,
\EN
where $\alpha_{\rm AD}=\tau_{\rm AD}\,\JJ\cdot\BB/\rho_0$ as the
AD $\alpha$ effect, and $\eta_{\rm AD}=\tau_{\rm AD}\vA^2$ is the
corresponding diffusive effect, where $\vA=|\BB|/\sqrt{\mu_0\rho_0}$
is the local Alfv\'en speed, although the variation of density is here
deliberately ignored in comparison with the actual Alfv\'en speed.

In addition to the usual kinetic to magnetic energy dissipation ratio,
\EQ
r_{\rm M}=\epsK/\epsM,
\EN
it is interesting to compute also the ratio of kinetic energy dissipation
to the sum of magnetic and AD dissipations,
\EQ
r_{\rm AD}=\epsK/(\epsM+\epsAD).
\EN
Likewise, in addition to the usual Prandtl number, $\Pm$, we also
quote the ambipolar Prandtl number, i.e.,
\EQ
\PAD=\nu/(\eta+\bra{\etaAD}).
\label{PAD}
\EN
It is unclear whether this quantity plays any role in characterizing
the kinetic to magnetic energy dissipation ratio.
We will therefore compare plots of this ratio as functions of both
$\Pm$ and $\PAD$.

\subsection{$E$ and $B$ mode polarization}

As an additional analysis tool, we compute the parity-even and
parity-odd linear polarization modes of the magnetic field,
$E$ and $B$, respectively.
They depend on the detailed physics causing polarized emission,
but for our purpose it will suffice to compute the intrinsic linear
complex polarization as
\EQ
Q+\ii U=-\epsilon\,(B_x+\ii B_y)^2
\label{QUBxBy}
\EN
for any arbitrarily chosen $xy$ plane.
Here, $Q(x,y)$ and $U(x,y)$ are the Stokes parameters characterizing
linear polarization, and $\epsilon$ is the polarized emissivity, which
will be assumed constant.
The difference between models with constant and $\BB$-dependent values
of $\epsilon$ turns out to be small \citep{BBKMR19}.

We then compute the Fourier transforms of $Q$ and $U$, indicated by a
tilde, e.g., $\tilde{Q}(k_x,k_y)=\int Q(x,y)\,e^{\ii\kk\cdot\xx}\dd^2\xx$,
where $\xx=(x,y)$ and $\kk=(k_x,k_y)$ are the position and wavevectors
in the $xy$ plane.
We then compute \citep{Kamion97,SZ97}
\begin{equation}
\tilde{E}+\ii\tilde{B}=(\hat{k}_x-\ii\hat{k}_y)^2(\tilde{Q}+\ii\tilde{U}),
\end{equation}
where $\hat{k}_x$ and $\hat{k}_y$ are the $x$ and $y$ components
of the planar unit vector $\hatkk=\kk/k$, and $k=(k_x^2+k_y^2)^{1/2}$.
We then transform $\tilde{E}$ and $\tilde{B}$ back into real space
to obtain $E(x,y)$ and $B(x,y)$ at a given position $z$.

Earlier work revealed a surprising difference in the statistics of $E$
and $B$ in that the probability density function (PDF) of $E$ is negatively
skewed, while that of $B$ is not.
However, not much is known about $E$ and $B$ mode polarizations for
different types of turbulence simulations.
Therefore, we also compute and compare the PDFs of $E$ and $B$ for
all the models presented in this paper.

\section{Results}

\subsection{Comparison between one and two fluid models}

Before presenting in detail the results obtained in the one-fluid
approximation, it is important to verify that those results can also be
obtained in the more complete two-fluid model.
Here we examine both one-dimensional and three-dimensional
two-fluid models.

\begin{figure*}\begin{center}
\includegraphics[width=\textwidth]{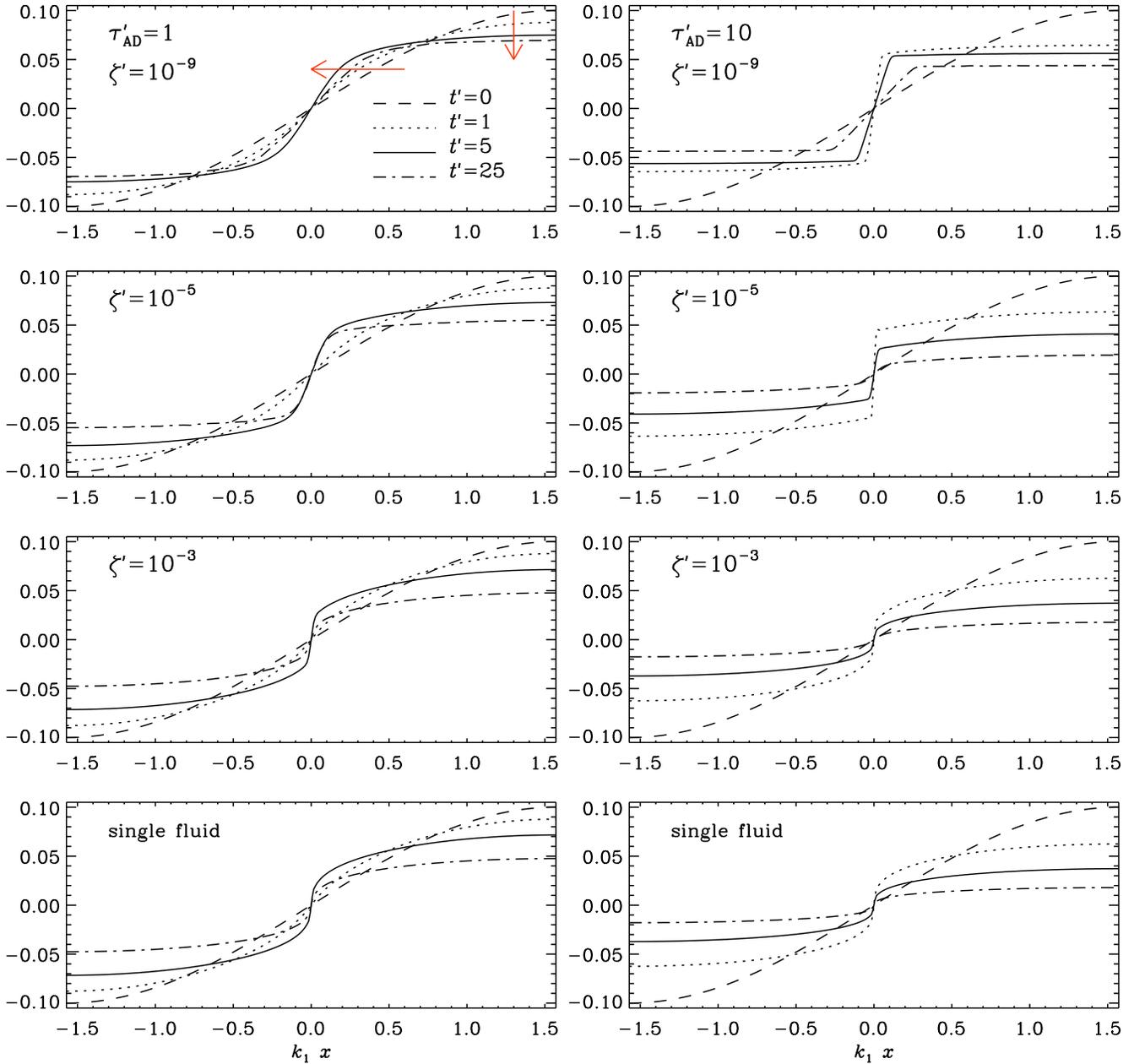}
\end{center}\caption[]{
Magnetic field profiles for $\tauAD'=1$ (left) and
$\tauAD'=10$ (right) with $\zeta'=10^{-9}$ (top),
$\zeta'=10^{-5}$ (second row), $\zeta'=10^{-3}$ (third row),
compared with magnetic field profile in the single fluid model (bottom).
The red arrows indicate the temporal evolution.
}\label{pcomp_var_8panels}\end{figure*}

\subsubsection{Formation of sharp structures in one dimension}

We examine here a two-fluid model similar to that of \cite{BZ95} to
demonstrates the similarity with the corresponding single fluid model.
As initial conditions, we choose for the magnetic field
$\BB=(0,\,B_0\sin k_1 x,\,0)$.
The $x$ component of the Lorentz force, $\partial B_y^2/2\partial x$
in this one-dimensional model, drives the charged fluid toward the
magnetic nulls at $x=0$ and $\pm\pi$.
If the resulting electron pressure gradient remains small enough,
this can lead to the formation of sharp structures.
In \Fig{pcomp_var_8panels}, we compare the results for three values
of $\zeta'$ and two values of $\rho_{\rm i0}/\rho_0$ ($10^{-3}$ and
$10^{-4}$) using $\gamma'=10^3$.
The two values of $\rho_{\rm i0}$ correspond to $\tauAD'=1$ and $10$,
respectively.
In all cases, we use $\alpha=\zeta\rho/\rho_{\rm i}^2$ to achieve initial
ionization equilibrium.
We choose $\Pm=20$, but used for $\eta k_1/\cs$ different values:
$10^{-4}$ for $\tauAD'=1$ and $2\times10^{-4}$ for $\tauAD'=10$,
while in all single fluid models we use $\eta k_1/\cs=5\times10^{-5}$.
We have increased $\nu$ and $\eta$ to avoid excessive sharpening of
the structures in our one-dimensional models.
We compare with the results from the one-fluid model in the last two
panels of \Fig{pcomp_var_8panels}.
We also compare models with $\tauAD'=1$ and $10$.

We see that for $\zeta'=10^{-3}$, good agreement between is the one-fluid
and two-fluid models is obtained.
The corresponding values of $\alpha$ for ionization equilibrium are
$10^3$ and $10^5$ for $\tauAD'=1$ and 10, respectively.
This encourages us to examines this model now in three dimensions.

\subsubsection{Spectral properties in three dimensions}

Next, we consider a setup similar to that studied below in more detail
in the one-fluid model.
Again, we consider the cases with $\tauAD'=1$ and $10$,
using $\zeta'=10^{-3}$, which was found to give good agreement with the
one-fluid model (cf.\ \Fig{pcomp_var_8panels}).
We consider here the case of relatively small magnetic diffusivity
($\eta k_1/\cs=5\times10^{-5}$), which will also be used in the one-fluid
models discussed below.

\begin{figure}\begin{center}
\includegraphics[width=\columnwidth]{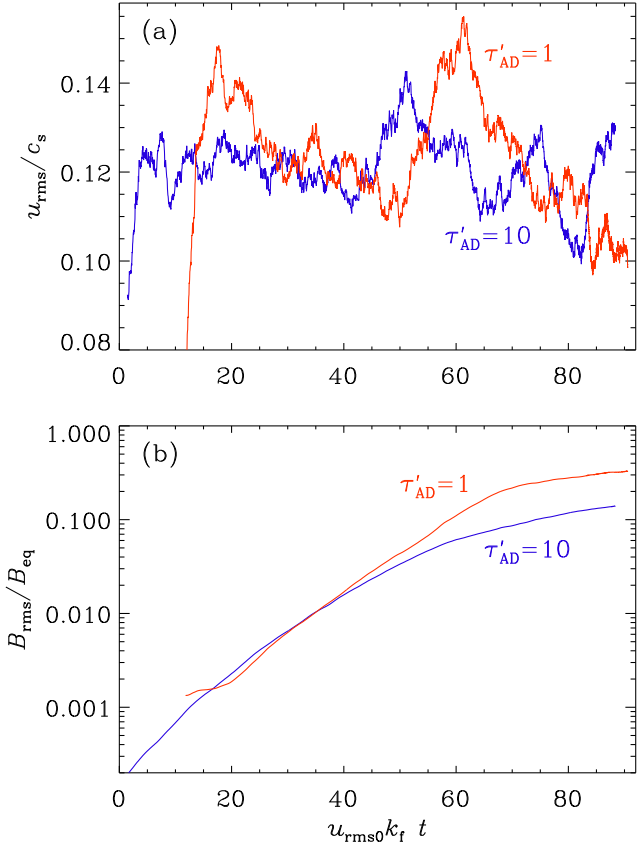}
\end{center}\caption[]{
(a) Evolution of the rms velocity (normalized by the sound speed)
for the runs with $\tauAD'=1$ (red) and $10$ (blue).
(b) Evolution of the rms magnetic field for the same runs.
}\label{p2FtauAD10and1_comp}\end{figure}

For both values of $\tau_{\rm AD}$, there is dynamo action with initial
exponential growth and subsequent saturation.
The mean instantaneous growth rate of the magnetic field, evaluated
by averaging $\lambda=\dd\Brms/\dd t$ over the duration of the early
exponential growth phase, is $\lambda/(\cs k_1)=0.019$.
In units of the turnover time, we have
$\lambda/(\urmsz\kf^{\rm eff})=0.080$.
For larger values of $\tau_{\rm AD}$, the dynamo saturates at a lower
magnetic field strength; see \Fig{p2FtauAD10and1_comp}.
Running the simulation beyond the early saturation shown here is
numerically expensive and would require higher resolution.
This is because of sharp gradients in the magnetic field.
This problem can be mitigated by increasing the viscosity of the ionized
fluid and certainly also by using a larger magnetic diffusivity, which
was also used in the one-dimensional runs shown in \Fig{pcomp_var_8panels}.
The dynamo would then become weaker, however, and this would no longer
be the model we would like to study in the one-fluid approximation below.

\begin{figure}\begin{center}
\includegraphics[width=\columnwidth]{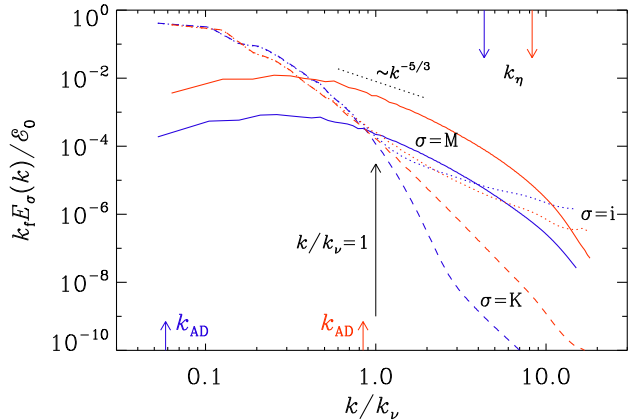}
\end{center}\caption[]{
Kinetic energy spectra for the neutral (dashed lines) and ionized fluids
(dotted lines) as well as magnetic energy spectra (solid lines) for
$\tau_{\rm AD}'=1$ (red) and $10$ (blue).
The $k^{-5/3}$ slope is shown for orientation.
}\label{pspecm_comp_neu}\end{figure}

In \Fig{pspecm_comp_neu}, we compare magnetic and kinetic energy spectra
for the two values of $\tau_{\rm AD}$.
They are normalized such that
\EQ
\int\EK(k)\,\dd k=\rho_0\bra{\uu^2}/2,\quad
\int\EM(k)\,\dd k=\bra{\BB^2}/2\mu_0.
\EN
Here, the kinetic energy is based on the neutral component, but we also
consider the kinetic energy of the ionized components, which we normalize
by the {\em same} density factor,
\EQ
\int\Ei(k)\,\dd k=\rho_0\bra{\uu_{\rm i}^2}/2.
\EN
This normalization has the advantage that we can more clearly see that
both velocity components are about equally big at large scales (small
$k$), when all spectra are also normalized by the same value, namely
the total kinetic energy of the neutrals, ${\cal E}_0=\rho_0\urms^2/2$.

We see that there is a marked separation between the ionized and neutral
fluid components for larger wavenumbers.
The wavenumber above which the two spectra diverge from each other is
independent of the value of $\tauAD$, and it is therefore also independent
of $k_{\rm AD}$, whose values are indicated by an arrow on the lower
abscissa of \Fig{pspecm_comp_neu}.
There is, however, a strikingly accurate agreement between the viscous
dissipation wavenumber, $k_\nu$, and the wavenumber where $\EK(k)$ and
$\Ei(k)$ begin to diverge from each other.
It therefore appears that the value of $k_{\rm AD}$ does not play
any role in the dynamics of turbulence with AD.
This confirms the earlier result of \cite{BS00} that the relevant
dissipation wavenumber is independent of AD and is just given
by the usual resistive wavenumber $k_\eta$, which was defined in
\Sec{EnergyDissipation} and agrees with the wavenumber defined by
\cite{XL16} after replacing $\epsM$ by $\kf\vA^3$.

We also see that the ionized fluid is not efficiently being dissipated
at the highest wavenumbers in this model: the kinetic energy spectrum
of the ionized fluid does not fall off as much as for the neutral fluid.
This is partially explained by the very low ion density in our model,
so the actual kinetic energy in the ionized fluid is still not very large.
Thus, the energy dissipation may appear insufficient because the amount
of energy to be dissipated is very small.

To understand why the magnetic field is apparently not visibly affected by
the breakdown of the strong coupling of the ionized and neutral species
below the viscous scale, we have to realize that for $\Pm=20\gg1$, the
velocity at $k\gg k_\nu$ is being driven entirely by the magnetic field.
Owing to the fact that $\rho_{\rm i}/\rho$ is very small 
($10^{-3}$ and $10^{-4}$ for $\tauAD'=1$ and $10$, respectively),
the velocity is too small to affect the magnetic field.
Instead, the magnetic field at large $k$ receives energy only from the
magnetic field at larger scales through a forward cascade.
This is also evidenced by the fact that, except for a vertical shift,
the magnetic spectrum looks similar for $\tauAD'=1$ and $10$.
This shows that the breakdown of the tight coupling below the resistive
scale will not affect our conclusions based on the single fluid
approximation considered in the main part of this paper.

\subsubsection{Conclusions from the two-fluid model}

We have seen that in the two-fluid model, the ionized and neutral
components are tightly coupled at large length scales ($k\ll k_\nu$).
At small scales, however, we see major departures between the two fluids.
There are clear differences in the results for the two values of
$\tau_{\rm AD}$ studied above.
For the larger value of $\tau_{\rm AD}$, the magnetic energy saturates
at a smaller value.
The magnetic field can therefore no longer drive turbulent motions beyond
the viscous cutoff scale, where $\EK(k)$ would normally fall off sharply
when there is no magnetic field.
For the ionized component, on the other hand, the difference between
the two spectra is much smaller and a comparatively high fraction of
kinetic energy still exists in the ionized component.
This is probably indicative of a significant fraction of small-scale
magnetic field structures where the ionized and neutral components
are counter-streaming in a way similar to what is seen in
\Fig{pcomp_var_8panels}.
After these preliminary studies, we now proceed with the examination of
the one-fluid model, which is simpler, but shows similar characteristics
and dependencies on $\tauAD'$, as we will see.

\subsection{The dynamo in one-fluid models}

\subsubsection{Kinematic evolution}

\begin{figure}\begin{center}
\includegraphics[width=\columnwidth]{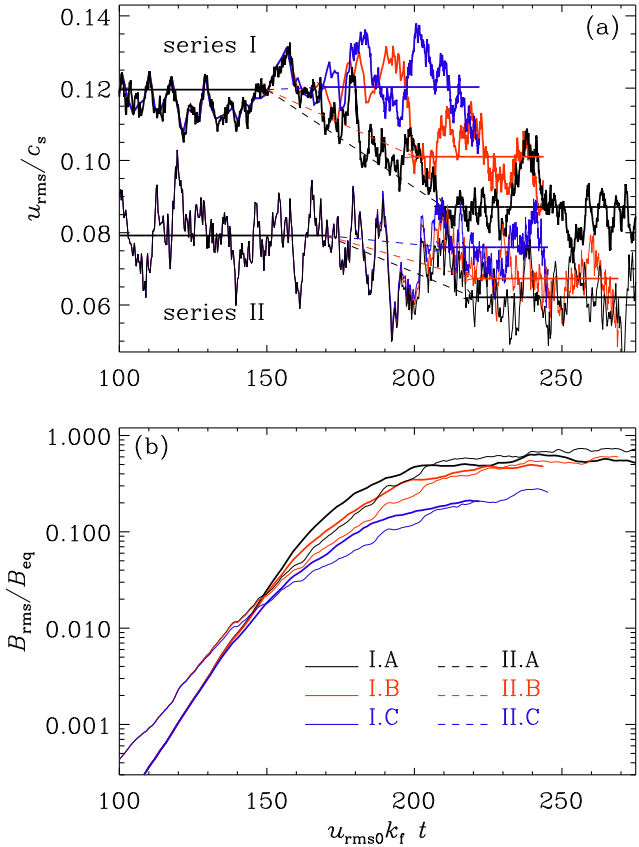}
\end{center}\caption[]{
(a) Evolution of the rms velocity (normalized by the sound speed)
for each of the three runs of series~I and II.
The values of late time averages are indicated by horizontal lines
in the corresponding color and connected by dashed arrows to the
corresponding horizontal line for the kinematic stage.
(b) Evolution of the rms magnetic field for series~I (solid lines)
and II (dashed lines) for small (black lines for runs~I.A and II.A),
intermediate (red lines for I.B and II.B), and large values (blue lines
for I.C and II.C) of $\StAD$.
}\label{pcomp}\end{figure}

Turning now to the study of dynamo action in the one-fluid model, we first
look at the evolution of the rms velocity and magnetic field versus time;
see \Fig{pcomp}.
The magnetic Reynolds numbers of the runs are 1200 for series~I and 790
for series~II.
This lower value for series~II is caused by the ten times larger viscosity
in this case ($\nu/\cs k_1=10^{-2}$ instead of $10^{-3}$).
We clearly see exponential growth in both cases.
The mean instantaneous growth rates are given by $\lambda/(\cs k_1)=0.019$
and $0.010$ for series~I and II, respectively.
In units of the turnover time, we have $\lambda/(\urmsz\kf^{\rm eff})
=0.080$ and $0.062$ for series~I and II, respectively.
These values are compatible with the relation $\lambda_0\Rmz^{1/2}$
with $\lambda_0\approx0.0023$; see also Fig.~3 of \cite{HBD04} as well
as Fig.~3 of \cite{Bra09}, were similar values of $\Rmz\approx1000$
were found and the $\Rmz^{1/2}$ scaling was demonstrated.

\begin{figure}\begin{center}
\includegraphics[width=\columnwidth]{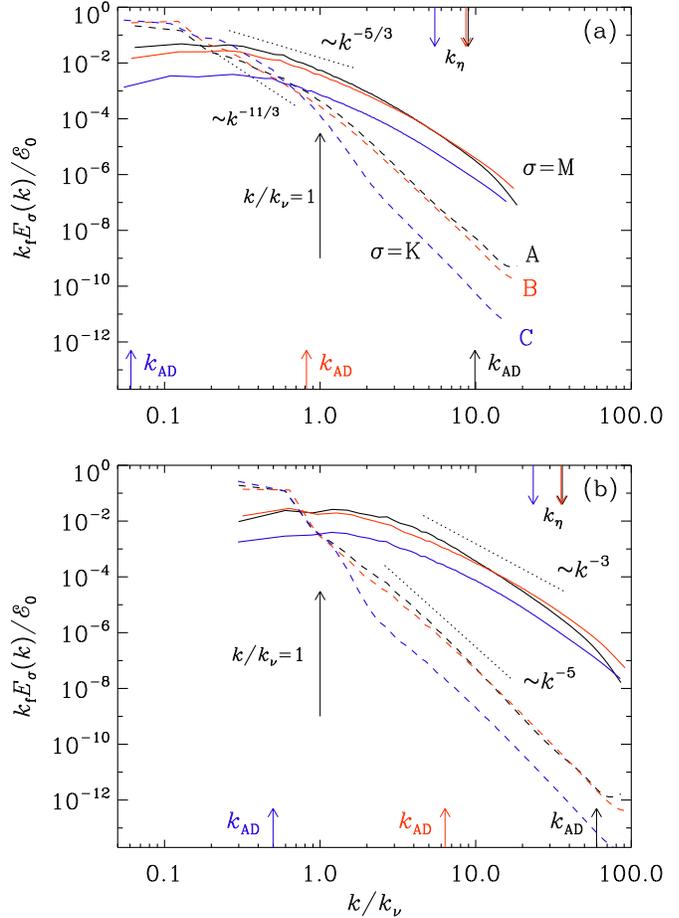}
\end{center}\caption[]{
Spectra of magnetic ($i={\rm M}$, solid lines) and kinetic
($i={\rm K}$, dashed lines) for each of the three runs in series~I
(top) and II (bottom).
}\label{pspecm_comp_both}\end{figure}

For all runs, the magnetic field eventually saturates owing
to the nonlinearity of the problem.
In addition to the  Lorentz force, $\JJ\times\BB$, there is the
AD nonlinearity.
It is a priori unclear which of the two is more important.
The saturation phenomenology of the small-scale dynamo has been
studied by \cite{Cho09}.
\cite{XL16} found that this dynamo saturation is independent of plasma
effects including AD.
Interestingly, \Fig{pcomp} now shows that for $\StAD\ge1$, the AD
nonlinearity does affect the solution, and this happens already when
$\Brms/\Beq\ge0.02$.
We also see that the kinetic energy decreases only very little during
saturation when AD is strong (cf.\ cases I.C and II.C).
This is because the velocity is only affected by the magnetic field,
whose saturation levels diminish with increasing values of $\StAD$.

\subsubsection{Spectral properties}

Next, we consider kinetic and magnetic energy spectra for series~I and II,
$\EK(k,t)$ and $\EM(k,t)$, respectively.
For both series, the kinetic energy spectra are found to be unaffected
for $k<k_\nu$, while the magnetic energy is clearly suppressed by AD at
all wavenumbers.
The magnetic energy spectrum does not really show power law scaling,
but it has a slope compatible with $k^{-5/3}$, although the spectrum
tends to become slightly shallower
at high wavenumbers when AD is strong (compare the red and blue lines
in \Fig{pspecm_comp_both} with the black ones).
This could be a signature of sharp structures that are expected to develop
in the presence of AD \citep{BZ94,ZB97}.
Sharp structures could be responsible for producing enhanced power at
high wavenumbers.
This is an effect that was also seen in the turbulence simulations
of \cite{BS00}.

\begin{figure}\begin{center}
\includegraphics[width=\columnwidth]{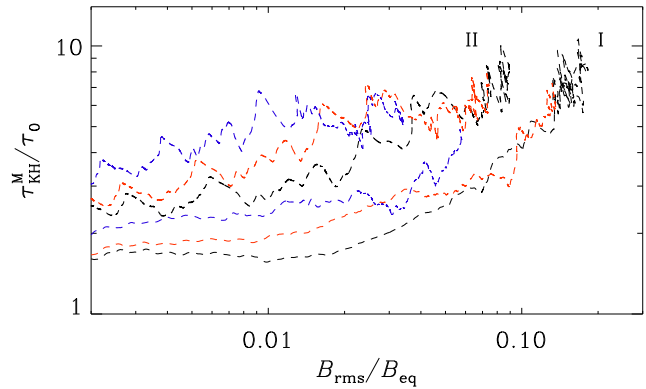}
\end{center}\caption[]{
Magnetic Kelvin-Helmholtz time normalized by the turnover time
versus normalized magnetic field strength.
}\label{pcomp_param}\end{figure}

In both series~I and II, the kinetic energy spectrum
develops a clear power law in the dissipation range, especially for
series~II, where power law scaling extends over about 1.5 decades, while
for series~I, the same power law is seen for only about half a decade.
The power law scaling of $\EK(k)$ is solely a consequence of magnetic
driving at $k>k_\nu$ when $\Pm$ is large.

\begin{figure*}\begin{center}
\includegraphics[width=\textwidth]{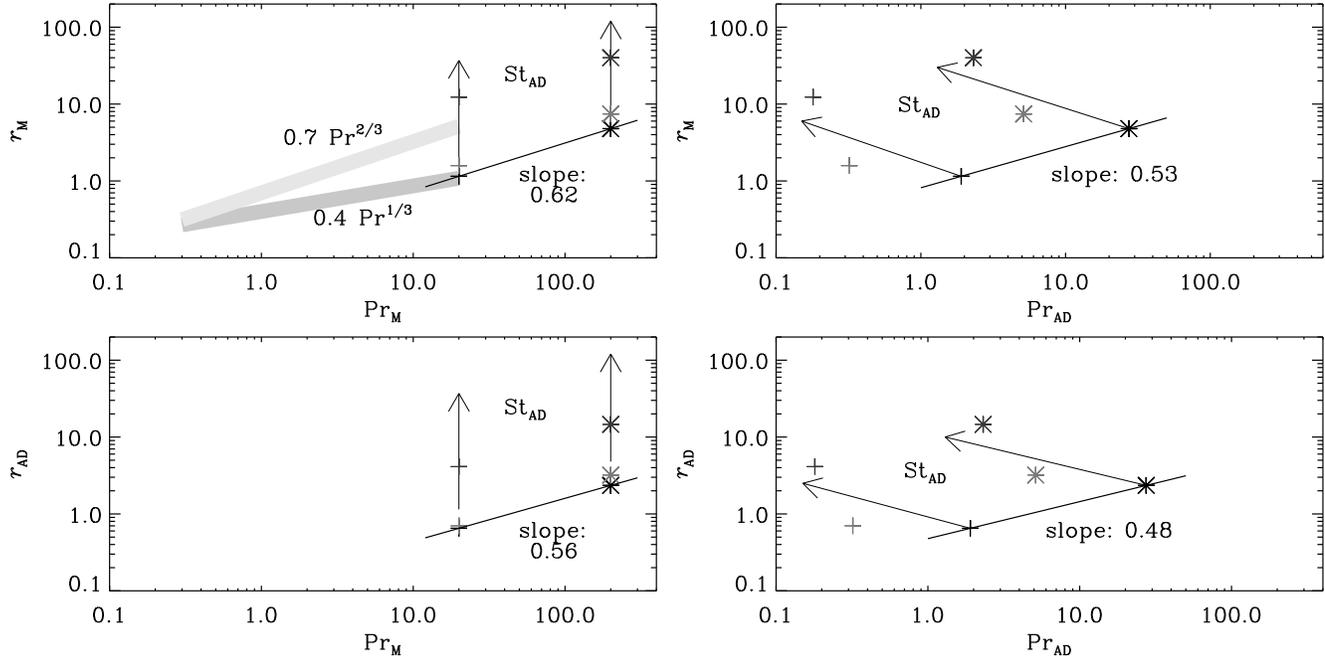}
\end{center}\caption[]{
Ratio of kinetic to magnetic and kinetic to ambipolar dissipation rates versus
magnetic and ambipolar Prandtl numbers.
The light and darker gray lines denote the scaling found by \cite{Bra14}
for large- and small-scale dynamos, respectively.
}\label{ptable}\end{figure*}

Also the magnetic energy spectrum shows a range with
power law scaling for series~II, where $\EM\propto k^{-5/3}$.
For series~I the $k^{-5/3}$ scaling is not so clear.
The kinetic energy spectrum is much steeper and has a slope
comparable with a $k^{-11/3}$ spectrum.
This is reminiscent of the Golitsyn spectrum of magnetic energy, which
applies to the opposite case of small magnetic Reynolds numbers \citep{Gol60}.
In that case, the electromotive force is balanced by the magnetic
diffusion term rather than the time derivative of $\BB$.
The similarity suggests that in the present case, the velocity is
driven through the balance between the Lorentz force and the viscous
force (which is proportional to $\nu\nabla^2\uu$) rather than through
a balance with the $\DD\uu/\DD t$ inertial term.

The magnetic energy spectrum peaks at a wavenumber $k_\ast$ that can
roughly be estimated by Subramanian's formula $k\ast\approx\kf\Rmc^{1/2}$
\citep{Sub99}.
Estimating $\Rmc\approx40$ for the critical magnetic Reynolds number for
dynamo action \citep{HBD04}, we have $k_\ast/k_\nu\approx0.5$ and $2.8$
for series~I and II, respectively.
This is in fair agreement with the position of the magnetic peak
wavenumber seen in \Fig{pspecm_comp_both}.  \cite{Schober15} proposed
a revised estimate with an exponent $3/4$ for Kolmogorov turbulence and
a larger prefactor, so the corresponding values are by about a factor
of eight larger.
I addition, both estimates would yield bigger values if $2\pi$ factors
in their definitions of $\Rm$ were taken into account.

\subsubsection{Comment on numerical diffusion}

At this point, a comment on the accuracy and properties of
the numerical scheme is in order.
The results presented above relating to the spectral kinetic energy
scaling in the high magnetic Prandtl number regime rely heavily upon
the presence of proper diffusion operators.
In fact, those are the only terms balancing an otherwise catastrophic
steepening of gradients by the $\uu\cdot\nab\uu$, $\uu\times\BB$, and
$\JJ\times\BB$ nonlinearities.
The weakly stabilizing properties of any third order time stepping scheme
and the dispersive errors of the spatial derivative operators such as
$\uu\cdot\nab$ do not contribute noticeably to numerical diffusion below
wavenumbers of half the Nyquist wavenumber \citep{Bra03}, which is the
largest wavenumber shown in our spectra.
This is different from codes that solve the ideal hydromagnetic equations.
Those codes prevent excessive steepening of gradients by the numerical
scheme in ways that cannot be quantified by an actual viscosity or
diffusivity.
This is sometimes also called numerical diffusion, but such a procedure
it is not invoked in the numerical simulations presented here.

\subsubsection{Magnetic dissipation}

If the magnetic field were not constantly regenerated by dynamo action,
it would decay on a timescale that we call the magnetic
Kelvin-Helmholtz time,
\EQ
\tau_{\rm KH}^{\rm M}=\EEM/\epsM.
\EN
In \Fig{pcomp_param}, we plot its instantaneous value versus the
instantaneous magnetic field strength as the dynamo saturates and the
field strength thus increases.
Almost independently of the presence or absence of AD
and regardless of whether we consider series~I or II, the ratio
$\tau_{\rm KH}^{\rm M}/\tau_0$ is always around eight; see the two
concentrations of data near $\Brms/\Beq\approx0.08$ and $0.16$ for series~I
and II, respectively.

\begin{figure*}\begin{center}
\includegraphics[width=\textwidth]{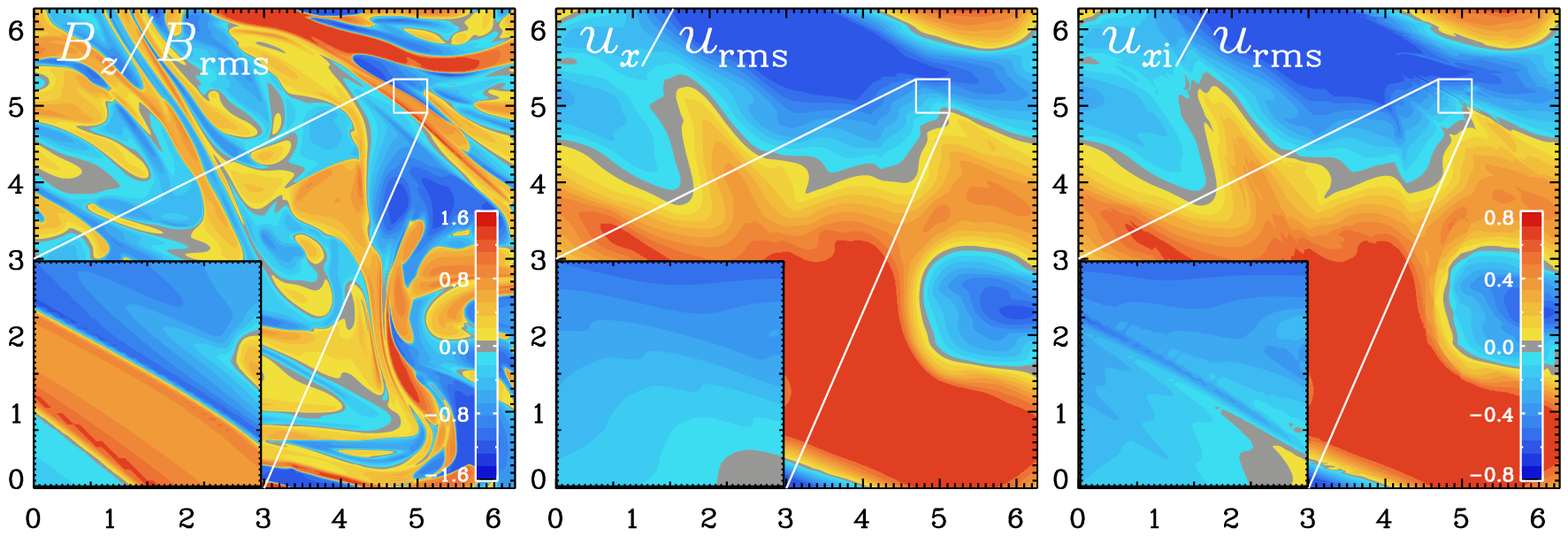}
\end{center}\caption[]{
Visualizations of $B_z/\Brms$, $u_x/\urms$, and $u_{x{\rm i}}/\urms$ for
the two-fluid model with $\StAD=0.15$ or $\tauAD'=1$.
The insets show a blow-up near a magnetic structure.
}\label{pslice_2F576Pm20AD1d}\end{figure*}

In the absence of AD, it was found that the ratio $r_{\rm M}=\epsM/\epsK$
of magnetic to kinetic energy dissipation increases with increasing
values of $\Pm$ like $\Pm^{1/3}$ for small-scale dynamo action and like
$\Pm^{2/3}$ for large-scale dynamo action (in the presence of
kinetic helicity of the turbulent flow).
In the presence of AD, there is an additional mode of dissipation
proportional to $\epsAD$.
On the other hand, also the effective magnetic Prandtl number is modified
if we include $\etaAD$ in the definition of $\Pm$, as in \Eq{PAD}.
The question is therefore whether there is any analogy between Ohmic
dissipation and dissipation through AD.
To assess this, we plot in \Fig{ptable} all four possibilities:
$r_{\rm M}$ versus $\Pm$ and $\PAD$, as well as $r_{\rm AD}$
versus $\Pm$ and $\PAD$.

Both $r_{\rm M}$ and $r_{\rm AD}$ are seen to increase with $\StAD$,
so the data points generally move upward in all four plots.
However, as we increase $\StAD$, we also decrease $\PAD$, so the data
points move to the left in \Fig{ptable}.
In this sense, there is no analogy with Ohmic dissipation.
It should be noted, of course, that both Ohmic dissipation and AD are
no longer accurate descriptions of the physics on small length scales.
It would therefore be interesting to revisit this question when
such an analysis of the full kinetic equations becomes feasible; see
\cite{Rincon16} and \cite{ZWUB17} for relevant references.
It is worth noting in this connection that the case with $\Pm\gg1$ is
special because the work done against the Lorentz force, which quantifies
the conversion of kinetic to magnetic energy, only operates on large
length scales when $\Pm\gg1$.
At small length scales, the sign of this term is reversed, so \cite{BR19}
called  this reversed dynamo action.
This means that the magnetic energy is not ohmically dissipated at small
length scales, but viscously.
\cite{BR19} speculated further that this loss of energy would really
correspond to the energization of ions and electrons, although there is
currently no evidence that this similarity is quantitatively accurate.

\begin{figure*}\begin{center}
\includegraphics[width=\textwidth]{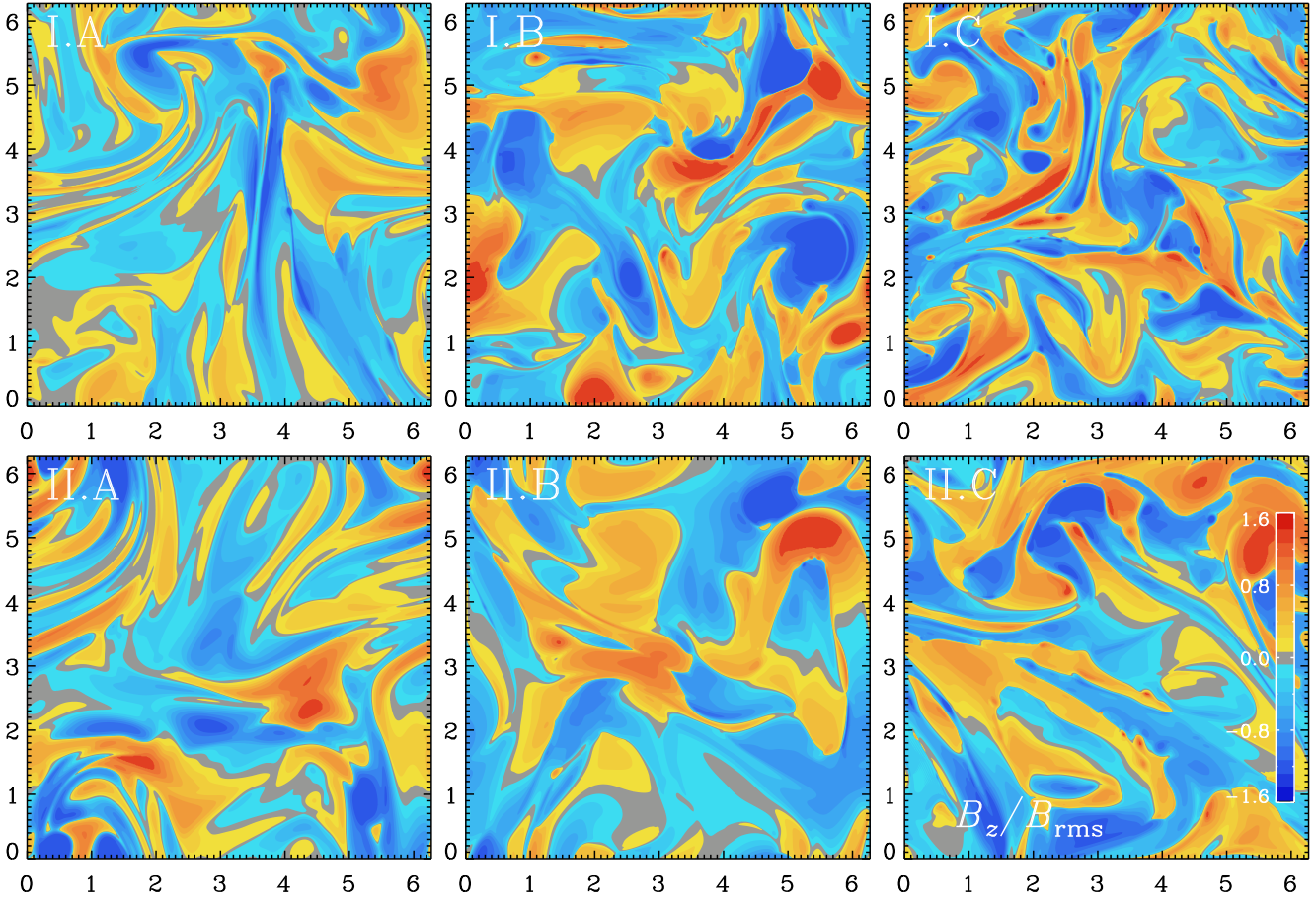}
\end{center}\caption[]{
Visualizations of $B_z(x,y)/\Brms$ for the single fluid models,
Runs~I.A--C and II.A--C.
}\label{pslice_panels}\end{figure*}

\begin{table*}\caption{Summary of the runs discussed in the paper.
}\vspace{12pt}\centerline{\begin{tabular}{crrcccccrrccc}
Run  & $\Rm$ & $\Pm\!\!$ & $\PAD$ & $r_{\rm M}$ & $r_{\rm AD}$
& $\St_{\rm AD}$ & $\!\!\bra{E^2}/\bra{B^2}\!\!$ 
& $\mbox{skew}\,E$ & $\mbox{skew}\,B$
& $\mbox{kurt}\,E$ & $\mbox{kurt}\,B$ & $\mbox{kurt}\,\BB$ \\
\hline
 I.a& 800& 20& 18.3 & 0.84 & 0.79 &0.00012&$1.66$&$ 2.05$ &$ 0.19$ &$14.9 $ &$3.38$& 2.33 \\
 I.b& 840& 20& 15.7 & 0.86 & 0.73 &0.00039&$1.80$&$ 2.00$ &$-0.36$ &$11.7 $ &$3.60$& 1.92 \\
 I.c& 850& 20& 10.5 & 0.97 & 0.71 &0.00130&$1.60$&$ 1.32$ &$ 0.04$ &$ 4.58$ &$1.20$& 1.35 \\
 I.d& 830& 20& 5.15 & 1.15 & 0.72 & 0.0038&$1.41$&$ 0.97$ &$-0.05$ &$ 4.99$ &$2.57$& 0.66 \\
 I.A& 860& 20& 1.9  & 1.15 & 0.65 & 0.013 &$1.46$&$ 0.85$ &$ 0.01$ &$ 6.30$ &$3.73$& 0.08 \\
 I.e& 800& 20& 0.71 & 1.25 & 0.65 & 0.037 &$1.33$&$ 0.41$ &$ 0.17$ &$ 5.34$ &$3.21$&$-0.06$\\
 I.B&1000& 20& 0.32 & 1.58 & 0.70 & 0.15  &$1.21$&$-0.18$ &$ 0.02$ &$ 1.77$ &$1.08$&$-0.43$\\
 I.C&1170& 20& 0.18 & 12.3 & 4.12 & 1.79  &$1.12$&$-0.27$ &$ 0.05$ &$ 2.08$ &$1.18$&$-0.57$\\
\hline
II.A& 630&200& 27.4 & 4.79 & 2.35 & 0.010&$1.27$&$ 0.72$ &$-0.13$ &$ 2.17$ &$1.32$& 3.19 \\
II.B& 670&200& 5.13 & 7.42 & 3.20 & 0.10 &$1.43$&$ 0.06$ &$-0.08$ &$ 0.91$ &$1.75$& 2.50 \\
II.C& 770&200& 2.31 & 40.0 & 14.6 & 1.19 &$1.29$&$-0.48$ &$-0.04$ &$ 3.71$ &$1.71$& 2.48 \\
\label{Tsummary}\end{tabular}}\end{table*}

\subsection{Spatial features related to AD}

\subsubsection{Visual inspection}

In \Fig{pslice_2F576Pm20AD1d}, we show $xy$ slices of $B_z/\Brms$
and compare with slices of the $x$ component of the neutral and ionized
flows, $u_x/\urms$ and $u_{x{\rm i}}/\urms$, respectively, in the same
(arbitrarily chosen) plane.
The magnetic field displays folded structures in places, as was first
emphasized by \cite{Scheko04}, but \cite{BS05} found that there are also
many other places in the volume that are not strongly folded.
Some of the folds lead to differences between the neutral and ionized
fluid components; see the insets of \Fig{pslice_2F576Pm20AD1d}.
In most other places, however, the two velocity species are remarkably
similar.
The $y$ and $z$ components of $\uu$ and $\uu_{\rm i}$ are also similar
to each other and show only small differences near magnetic structures.

Next, we compare the magnetic field for different one-fluid models; see
\Fig{pslice_panels}, where we compare the three models of series I and II.
The overall magnetic field strength is weaker for model C compared with
models B and A.
To remove this aspect from the comparison, we plot in \Fig{pslice_panels}
the $B_z$ components of the magnetic field normalized by the rms values
for each model.

It is hard to see systematic differences between the different cases.
There could be more locations with strong horizontal gradients in
$B_z(x,y)$, where $\StAD$ is large (compare Runs~C of series~I and II
with Runs~A and B of the corresponding series), but the resulting changes
are not very obvious.
There are also no clear differences between series~I and II themselves.
For these reasons, it is important to look at statistical measures to
study the differences.
This will be done next.

\subsubsection{Statistical analysis}

\begin{figure}\begin{center}
\includegraphics[width=\columnwidth]{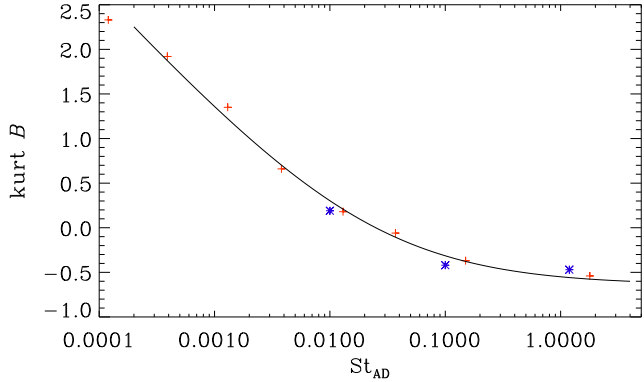}
\end{center}\caption[]{
Dependence of $\mbox{kurt}\,\BB$ on $\St_{\rm AD}$.
The red (blue) symbols denote the results for series~I (II).
}\label{pkurt}\end{figure}

In this section, we investigate in more quantitative detail the effects
of AD on the structure of the magnetic field.
We know that AD tends to clip the peaks of the magnetic field
at locations where its strength is large \citep{BZ95}.
This should lead to a reduced kurtosis,
\EQ
\mbox{kurt}\,B_i=\bra{B_i^4}/\bra{B_i^2}^2-3.
\EN
It is unclear, however, whether this is a statistically significant effect.
To examine this, we compute the resulting values of $\mbox{kurt}(B_i)$.
Since our simulations are isotropic, we can improve the statistics
of the kurtosis by taking the average over all three directions, i.e.,
we define $\mbox{kurt}\,\BB$ (bold without subscript on $\BB$) as
\EQ
\mbox{kurt}\,\BB=(\mbox{kurt}\,B_x+\mbox{kurt}\,B_y+\mbox{kurt}\,B_z)/3,
\EN
and compute it for each of the two series and for
different values of $\St_{\rm AD}$.
In this context, we recall that the kurtosis vanishes for gaussian-distributed
data, and it is 3 for an exponential distribution.
Here we find a systematic crossover from values somewhat smaller than 3
to negative values when $\StAD\ga0.02$; see \Fig{pkurt} for series~I and
II with $\Pm=20$ and $200$, respectively.
Here we have included the additional runs~I.a--e with lower values $\StAD$
have been added.
This dependence can roughly be described by a fit of the form
\EQ
\ln\mbox{kurt}\,\BB=e^{\kappa_\infty}+\StAD^{-\alpha},
\EN
where $\kappa_\infty\approx2.36$ is the value of $\mbox{kurt}\,\BB+3$
for large values of $\StAD$ and $\alpha\approx0.61$ is the slope for
smaller values.
Additional terms and parameters could be included in this fit to
account for finite values of the kurtosis for $\StAD\to0$, but this
does not appear to be necessary for describing the present data; see
\Tab{Tsummary}.
In conclusion, it appears that the measurement of the kurtosis of the
magnetic field in the interstellar medium could be a useful diagnostic
tool that should be explored further in future.

\begin{figure}\begin{center}
\includegraphics[width=\columnwidth]{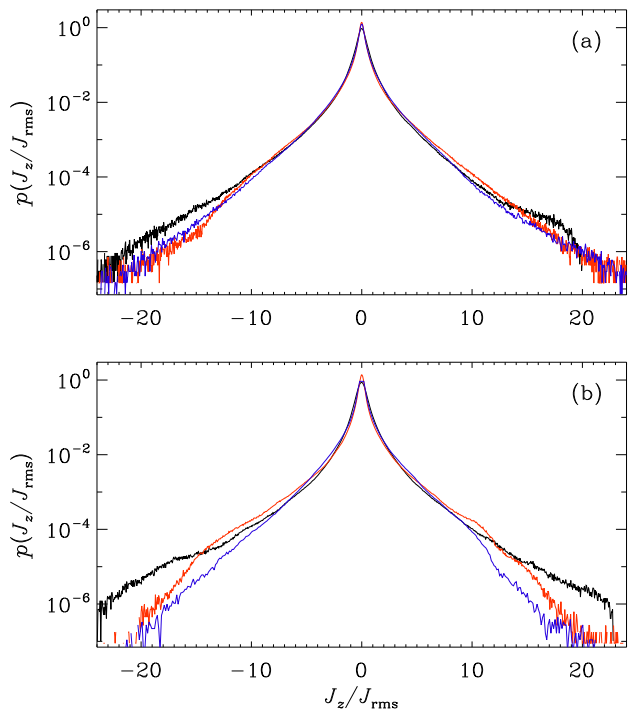}
\end{center}\caption[]{
Histograms of $J_z$ for (a) series~I and (b) series~II.
Black, red, and blue lines denote the cases A, B, and C,
respectively.
}\label{phisto_jz_comp}\end{figure}

In \Fig{phisto_jz_comp} we show histograms of $J_z$ for series~I and II.
We see that, as $\StAD$ is increased, the wings of the distributions are
being clipped slightly.
On the other hand, the amount of clipping is actually relatively
small compared with the increase in magnetic field strength as
$\StAD$ is increased.
This is to be expected, because AD tends to create force-free regions
where $(\JJ\times\BB)^2$ is minimized and $(\JJ\cdot\BB)^2$ is maximized.
In between those regions, on the other hand, there are sharp current
sheets that were already found in the earlier work of \cite{BZ94}.

It is important to note that one usually never measures the
magnetic field directly, but instead the linear polarization through
either synchrotron radiation or through dust emission.
In both cases, it therefore appears useful to discuss the two rotationally
invariant modes of linear polarization, namely the $E$ and $B$ mode
polarizations.
This will be done in the next section.

\begin{figure*}\begin{center}
\includegraphics[width=\textwidth]{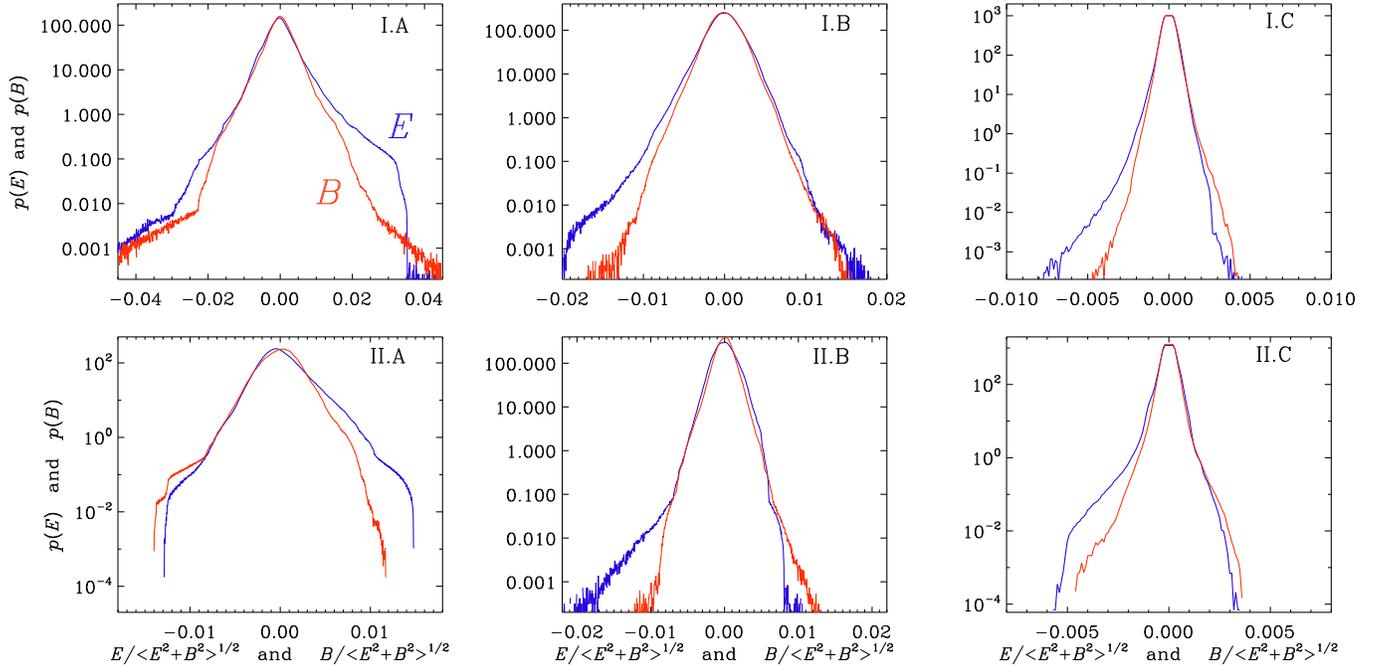}
\end{center}\caption[]{
$E$ and $B$ mode polarizations for series~I (upper row) and II (lower row).
Blue (red) lines denote the normalized probability density functions of
$E$ ($B$) mode polarization.
}\label{phisto_eb_comp}\end{figure*}

\begin{figure*}\begin{center}
\includegraphics[width=\textwidth]{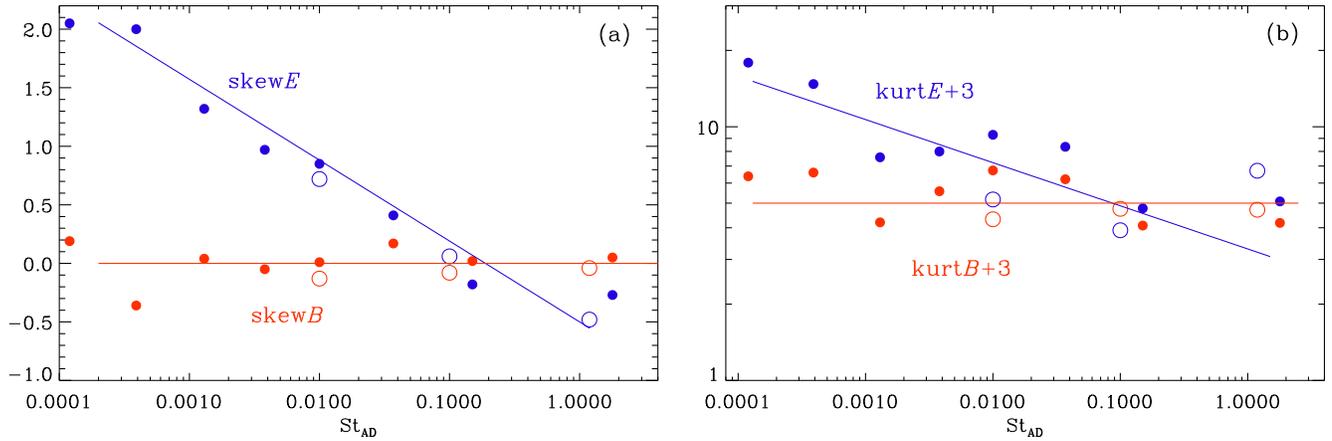}
\end{center}\caption[]{
(a) Dependence of $\mbox{skew}\,E$ (blue) and $\mbox{skew}\,B$ (red) on $\St_{\rm AD}$
and (b) dependence of $\mbox{kurt}\,E+3$ (blue) and $\mbox{kurt}\,B+3$ (red) on
$\St_{\rm AD}$.
Filled (open) symbols refer to series~I (II).
The straight lines represent approximate fits given by
$\mbox{skew}\,E=-0.5-0.3\ln\StAD$ (blue) and
$\mbox{skew}\,B=0$ (red) in (a), and
$\mbox{kurt}\,E+3=3.3\,\StAD^{0.17}$ (blue) and
$\mbox{kurt}\,B=2$ (red) in (b).
}\label{ptable2}\end{figure*}

\subsection{$E$ and $B$ mode polarizations}

The analysis of $E$ and $B$ mode polarization has been particularly
important in the context of cosmology \citep{Kamion97,SZ97} and, more
recently, in the context of dust foreground polarization \citep{Adam16}.
It was found that there is a systematic excess of $E$ mode power over
$B$ mode power by about a factor of two, which was unexpected at the
time \citep{CHK17}.
Different proposals exist for the interpretation of this.
It is possible that the excess of $E$ mode polarization is primarily
an effect of the dominance of the magnetic field, i.e., a result of
magnetically over kinetically dominated turbulence \citep{KLP17}.
Using simulations of supersonic hydromagnetic turbulent star formation,
\cite{Kritsuk18} found that the observed $E$ over $B$ ratio can be
reproduced.
However, not enough work has been done to assess the full range of
possibilities for different types of flows.
For solar linear polarization, for example, it has been found that
there is no excess of $E$ over $B$ mode polarization, although the
possibility of instrumental effects has not yet been conclusively
addressed \citep{BBKMR19}.

Looking at \Fig{phisto_eb_comp}, we see that, as $\StAD$ is increased,
there is a systematic change of the skewness of $E$ (but not of $B$)
as $\StAD$ is increased.
For small values of $\StAD$, the skewness is positive and for large
values it is negative.
Here we define the skewness as
\EQ
\SKEW E=\bra{E^3}/\sigma_E^3,\quad
\SKEW B=\bra{B^3}/\sigma_B^3,
\EN
where $\sigma_E^2=\bra{E^2}-\bra{E}^2$ and
$\sigma_B^2=\bra{B^2}-\bra{B}^2$ are their variances.
Note that here the $B$ is not to be confused with
the components $B_i$ of the magnetic field, which are related to each other
only through \Eq{QUBxBy}.

The increase of the skewness of $E$ with $\StAD$ is seen both for series~I
(where $\mbox{skew}\,E=-0.27$ for $\StAD\approx1.8$ in I.C) and series~II
(where $\mbox{skew}\,E=-0.48$ for $\StAD\approx1.2$ in II.C).
For small values of $\StAD$, however, there is a much more dramatic effect
in that $\mbox{skew}\,E$ reaches values of around $2$, which is much more
extreme than what was found earlier for decaying hydromagnetic turbulence.
Even a change of $\StAD$ from $10^{-2}$ (I.A) to $10^{-4}$ (II.a),
has a strong effect in that $\mbox{skew}$ changes from 0.85 to 2.
The kurtosis of $E$ reaches more extreme values much larger than 10;
see \Fig{Tsummary} for a summary of the statistics of $E$ and $B$.
Although we have not determined error bars, we can get a sense of
the reliability of the data by noting that the trend with $\StAD$
is reasonably systematic; see \Fig{ptable2}.

In view of the negative skewness found previously for decaying
hydromagnetic turbulence \citep{BBKMR19}, it now appears that negative
skewness of $E$ is not a general property of hydromagnetic turbulence,
although it may well appear in the interstellar medium where both AD
can be present and magnetic fields can be significant.
AD can also play a role in the solar chromosphere, where it contributes
to heating cold pockets of gas \citep{KC12}.
It needs to be checked whether this can lead to observable effects.
The analysis of $E$ and $B$ mode polarization is therefore, an interesting
diagnostic tool, although more work needs to be done to learn about all
the possible ways of interpreting those two modes of polarization.

\section{Conclusions}

In the cold interstellar medium, ionization and recombination are important.
The electron pressure can then be neglected and the single fluid approach
of AD becomes an excellent approximation.
Our work has now demonstrated that AD does not have diffusive
properties in the sense of enhancing the effects of microphysical
magnetic diffusion.
This is most likely due to the fact that AD is a nonlinear effect
that operates only in places where the field is strong in the sense
that $\tauAD\vA^2\gg\tau_0\urms^2$.
In fact, in one dimension it is easy to see that the Lorentz force
acting on the ionized fluid works in such a way as to move
more ionized fluid towards the magnetic null \citep{BZ95}.
This depletes the field maxima and leads to a pile-up of
magnetic field just before the magnetic null.
This effect is particularly pronounced when $\tauAD\gg\tau_0$,
and thus $\StAD\gg1$.

Although the spectral shape at large $k$ is only weakly affected by AD,
it does have a clear effect on the kinetic energy spectrum at $k>k_\nu$
and suppresses the spectral kinetic energy of the neutrals markedly.
The kinetic energy of the charged species is even slightly enhanced.
This is surprising, because the overall rms velocity of the neutrals is
hardly affected at all.
One must keep in mind, however, that not much kinetic energy is
contained deep in the kinetic energy tail at large $k$.
In fact, the only reason why there is some level of kinetic energy at
all is that, owing to the large magnetic Prandtl number, there is still
significant magnetic energy at those high wavenumbers that drives the
kinetic motions.

From an observational point of view, we can identify two potentially
useful ways of diagnosing the importance of AD in the interstellar medium.
First, there is the direct effect on the statistics of the magnetic field.
The importance of AD can then potentially be quantified by measuring the
kurtosis of the components of the magnetic field.
Alternatively, there appears to be a systematic effect on the statistics
of the $E$ and $B$ mode polarizations.
While the $B$ mode polarization is generally unaffected by turbulence,
the $E$ mode polarization can exhibit non-vanishing skewness,
which is positive for a weak AD and negative for strong AD.
This is an unexpected signature in view of recent results for decaying
hydromagnetic turbulence, where the skewness was found to be negative
even without AD.

In this work, we have studied only two values of the magnetic
Prandtl number.
However, the effect of changing the value of $\Pm$ on observational
properties such as $E$ and $B$ is rather weak; see \Fig{ptable2}.
This is interesting because in cold molecular clouds, the magnetic
Prandtl number can potentially drop below unity.
It would therefore in future be useful to study whether the present
results carry over into the regime of lower values of $\Pm$ (possibly
below unity), and whether the effects on the skewness of $E$ and $B$
mode polarizations remain unchanged.

\section*{Acknowledgements}
I am grateful to Dinshaw Balsara, Alex Lazarian, and Siyao Xu
for useful comments.
I acknowledge the suggestions made by the referee to compare with
the more complete two-fluid description of AD.
I also thank Wlad Lyra for having implemented the two-fluid module
for AD in the {\sc Pencil Code}, and Ellen Zweibel for having
taught me all I know about AD.
This work was supported through
the National Science Foundation, grant AAG-1615100,
the University of Colorado through its support of the
George Ellery Hale visiting faculty appointment,
and the grant ``Bottlenecks for particle growth in turbulent aerosols''
from the Knut and Alice Wallenberg Foundation, Dnr.\ KAW 2014.0048.
The simulations were performed using resources provided by
the Swedish National Infrastructure for Computing (SNIC)
at the Royal Institute of Technology in Stockholm and
Chalmers Centre for Computational Science and Engineering (C3SE).


\vfill\bigskip\noindent\tiny\begin{verbatim}
$Header: /var/cvs/brandenb/tex/mhd/AD/paper.tex,v 1.98 2019/05/29 02:30:00 brandenb Exp $
\end{verbatim}

\end{document}